\documentclass[fleqn,twoside]{article}
\usepackage{amsfonts,bezier,epsf}

\def\noi{\noindent}

\makeatletter
\renewcommand{\section}{\@startsection{section}{1}{0pt}%
        {-3.5ex plus -1ex minus -.2ex}{2.3ex plus .2ex}%
        {\large\bf\protect\raggedright}}

\renewcommand{\subsection}{\@startsection{subsection}{2}{0pt}%
        {-3ex plus -1ex minus -.2ex}{1.4ex plus .2ex}%
        {\normalsize\bf\protect\raggedright}}

\renewcommand{\thesubsubsection}%
        {\arabic{section}.\arabic{subsection}.\arabic{subsubsection}.}
\newcommand{\para}{\@startsection{paragraph}{4}{0pt}%
        {1.5ex plus -.5ex minus -.2ex}{-1em}{\normalsize\bf}}

\renewcommand{\@oddhead}{\raisebox{0pt}[\headheight][0pt]{%
   \vbox{\hbox to\textwidth{\rightmark \hfil \rm \thepage \strut}\hrule}}}
\renewcommand{\@evenhead}{\raisebox{0pt}[\headheight][0pt]{%
   \vbox{\hbox to\textwidth{\thepage \hfil \leftmark \strut}\hrule}}}

\newcommand{\heads}[2]{\markboth{\protect\small\it #1}{\protect\small\it #2}}

\makeatother

\newcommand{\Title}[1]{\noi {\Large #1} \\}
\newcommand{\Author}[2]{\noi{\large\bf #1}\\[2ex]\noindent{\it #2}\\}

\newcommand{\Abstract}[1]{\vskip 2mm \begin{center}
        \parbox{16.4cm}{\small\noi #1} \end{center}\medskip}

\newcommand{\foom}[1]{\protect\footnotemark[#1]}

\newcommand{\email}[2]{\footnotetext[#1]{e-mail: #2}}

\newcommand{\Ref}[1]{Ref.\,\cite{#1}}
\newcommand{\sect}[1]{Sec.\,#1}

\def\nq{\hspace*{-1em}}
\def\nqq{\hspace*{-2em}}
\def\nhq{\hspace*{-0.5em}}

\def\cm{\hspace*{1cm}}
\def\inch{\hspace*{1in}}
\def\para{\paragraph}
\newcommand{\Theorem}[2]{\medskip\noi {\bf #1. \ }{\it #2}\medskip}

\newcommand{\Picture}[3]{
	\begin{figure} 	\centering \unitlength=1mm
	\begin{picture}(84,#1)
		\put(0,0){\line(0,1){#1}}            
		\put(0,0){\line(1,0){84}}
		\put(84,0){\line(0,1){#1}}
		\put(0,#1){\line(1,0){84}}
	\put(0,0){#2}                       \end{picture}
        \caption{\protect\small #3}  \smallskip \hrule \end{figure}
	}

\def\eq{Eq.\,}
\def\eqs{Eqs.\,}
\def\beq{\begin{equation}}
\def\eeq{\end{equation}}
\def\bear{\begin{eqnarray}}
\def\al{&\nhq}
\def\lal{&&\nqq {}}               
\def\bearr{\bear \lal}
\def\ear{\end{eqnarray}}

\def\tst{\textstyle}

\def\nn{\nonumber\\ {}}

\def\nnn{\nonumber\\ \lal }

\def\yy{\\[5pt] {}}

\def\eql{\al =\al}

\def\eqdef{\stackrel{\rm def}=}
\def\e{{\,\rm e}}
\def\d{\partial}

\def\sign{\mathop{\rm sign}\nolimits}

\def\const{{\rm const}}

\def\half{{\tst\frac{1}{2}}}

\def\Jl#1#2{{\it #1\/} {\bf #2},\ }

\def\CQG#1 {\Jl{Class. Qu. Grav.}{#1}}
\def\DAN#1 {\Jl{Dokl. AN SSSR}{#1}}
\def\GC#1 {\Jl{Grav. \& Cosmol.}{#1}}
\def\GRG#1 {\Jl{Gen. Rel. Grav.}{#1}}
\def\JETF#1 {\Jl{Zh. Eksp. Teor. Fiz.}{#1}}
\def\JMP#1 {\Jl{J. Math. Phys.}{#1}}
\def\NPB#1 {\Jl{Nucl. Phys.}{B\ #1}}
\def\PLA#1 {\Jl{Phys. Lett.}{#1A}}
\def\PLB#1 {\Jl{Phys. Lett.}{#1B}}
\def\PRD#1 {\Jl{Phys. Rev.}{D\ #1}}
\def\PRL#1 {\Jl{Phys. Rev. Lett.}{#1}}

\def\GR{general relativity}
\def\sph{spherically symmetric}
\def\ssph{static, spherically symmetric}
\def\fig{Fig.\,}
\def\bh{black hole}
\def\Sch{Schwarzschild}
\def\dS{de Sitter}

\def\mn{_{\mu\nu}}
\def\MN{^{\mu\nu}}
\def\mN{_\mu^\nu}

\def\oA{{\overline A}}
\def\od{{\overline d}}
\def\og{{\overline g}}

\def\M{{\mathbb M}}
\def\R{{\mathbb R}}
\def\S{{\mathbb S}}
\def\oM{{\overline \M}}

\heads{K.A. Bronnikov}
	{Scalar vacuum structure in general
	relativity and alternative theories. Conformal continuations}

\begin{document}
\twocolumn[
\thispagestyle{empty}

\rightline{\large\bf gr-qc/0110125}
\bigskip

\Title	{\bf Scalar vacuum structure in general
	relativity and alternative theories. \yy
	Conformal continuations}

\Author{K.A. Bronnikov\foom 1}
{Centre for Gravitation and Fundamental. Metrology, VNIIMS,
        3-1 M. Ulyanovoy St., Moscow 117313, Russia;\\
Institute of Gravitation and Cosmology, PFUR,
        6 Miklukho-Maklaya St., Moscow 117198, Russia}

\Abstract
    {We discuss the global properties of \ssph\ configurations of a
     self-gravitating real scalar field $\varphi$ in general relativity
     (GR), scalar-tensor and high-order (curvature-nonlinear) theories of
     gravity in various dimensions. In GR, for scalar fields with an
     arbitrary potential $V(\varphi)$, not necessarily positive-definite, it
     is shown that the list of all possible types of space-time causal
     structure in the models under consideration is the same as the one for
     $\varphi = \const$. In particular, there are no regular black holes
     with any asymptotics. These features are extended to scalar-tensor and
     curvature-nonlinear gravity, connected with GR by conformal mappings,
     unless there is a conformal continuation, i.e., a case when a
     singularity in a solution of GR maps to a regular surface in an
     alternative theory, and the solution is continued through such a
     surface. Such an effect is exemplified by exact solutions in GR with a
     massless conformal scalar field, considered as a special scalar-tensor
     theory. Necessary conditions are found for the existence of a conformal
     continuation; they only hold for special choices of scalar-tensor and
     high-order theories of gravity.  }

] 

\email 1 {kb@rgs.mccme.ru}

\section {Introduction}

   This paper continues the study of global properties of scalar-vacuum
   configurations in \GR, described by the action
\beq 						             \label{act-d}
    S = \int d^D x \,\sqrt{|g|} [R + (\d\varphi)^2 - 2V(\varphi)]
\eeq
   and similar systems in some alternative theories of gravity, begun in
   Refs.\,\cite{vac1,vac2}. Here $D$ is the number of space-time
   dimensions, $R$ is the scalar curvature, $g=\det (g\mn)$, $\varphi$ is a
   real scalar field, $(\d\varphi)^2 = g\MN\d_\mu\varphi\d_\nu\varphi$, and
   the function $V(\varphi)$ is a potential. This action in case $D=4$,
   with many particular forms of $V(\varphi)$, is conventionally used to
   describe the vacuum (sometimes interpreted as a variable cosmological
   term) in inflationary cosmology, for the description of growing vacuum
   bubbles, etc, to say nothing of diverse field-theoretical studies of
   scalar fields with different potentials. Where and when $\varphi=\const$,
   the potential $V(\varphi)$ behaves as a cosmological constant.

   In the latter case, \sph\ solutions to the Einstein equations
   (the \Sch-(anti-)\dS\ metric and its multidimensional extension)
   and their global properties in different special cases are well-known
\cite{SdS},
   see also \sect 2 of the present paper. The solutions are static due to the
   extended Birkhoff theorem (see \cite{bm-birk} and references therein), and
   all of them, except the solutions with zero mass parameter $m$, contain
   curvature singularities at the centre.

   A wider set of space-times is connected with the so-called false vacuum,
   i.e., the system with the action (\ref{act-d}). One might expect that the
   inclusion of scalar fields with various potentials should considerably
   increase the choice of possible qualitative behaviours of \ssph\
   configurations. There are, however, very strong general restrictions that
   follow directly from the field equations due to (\ref{act-d}).
   Thus, if $V\geq 0$, the only asymptotically flat BH solution in 4
   dimensions is Schwarzschild, as follows from the well-known no-hair
   theorems (see \Ref {bek98} for a recent review).  Another result concerns
   solitonic (particle-like) configurations with a regular centre and a flat
   asymptotic: if $V\geq 0$, then such a configuration cannot have a
   positive mass \cite{brsh}.

   It is of interest what can happen if the asymptotic flatness
   and/or $V\geq 0$ assumptions are abandoned. Both assumptions are
   frequently violated in modern studies. Negative potential energy
   densities, in particular, the cosmological constant $V=\Lambda < 0$
   giving rise to the anti--de Sitter (AdS) solution or AdS asymptotic, do
   not lead to catastrophes (if restricted below), are often treated in
   various aspects and quite readily appear from quantum effects like vacuum
   polarization.

   Our previous papers \cite{vac1,vac2} have provided some essential
   restrictions on the possible behaviour of solutions of the theory
   (\ref{act-d}) with arbitrary $V(\varphi)$ in $D$ dimensions. It has
   been shown that, whatever is the potential, the variable scalar field
   adds nothing to the list of causal structures known for $\varphi=\const$.
   The possibility of regular configurations without a centre (wormholes and
   horns) was also ruled out. Extensions of these results to some more
   general field models were indicated. Considered were (i) generalized
   scalar field Lagrangians in GR, with an arbitrary dependence on the
   $\varphi$ field and its gradient squared; (ii) multiscalar theories of
   sigma-model type; (iii) scalar-tensor theories (STT) of gravity; (iv)
   curvature-nonlinear (high-order) gravity (HOG) with the Lagrangian of the
   form $f(R)$ where $f$ is an arbitrary function. In items (iii) and (iv),
   conformal mappings are used to reduce the original field equations to
   those following from (\ref{act-d}).

   This paper pays special attention to the nature of these conformal
   mappings. The point is that, when a manifold $\M[g\mn]$ is conformally
   mapped to another manifold $\oM[\og\mn]$ (so that $\og\mn=F(x)\og\mn$),
   the global properties of both manifolds are the same as long as the
   conformal factor $F$ is everywhere smooth and finite. It can happen,
   however, that a singular surface in $\oM$ maps to a regular surface
   in $\M$ due to a singularity in the conformal factor $F$. Then $\M$
   can be continued in a regular manner through this surface, and the global
   properties if $\M$ can be considerably richer than those of $\oM$: in
   the new region, one can possibly find, e.g., new horizons or another
   spatial infinity. A known example of this phenomenon, to be called {\sl
   conformal continuation,\/} is provided by the properties of the \ssph\
   solution for a conformally coupled scalar field in GR \cite{bbm70,br73}
   as compared with the corresponding solution for a minimally coupled
   scalar field --- see \sect 6.

   It will be further shown that the mappings that connect STT and HOG
   (the so-called Jordan conformal frame) with GR with a minimally coupled
   scalar field described by the action (\ref{act-d}) (the Einstein frame),
   provide conformal continuations only under certain special requirements
   upon the original theory. Under very general conditions, conformal
   continuations are absent, and the global structure restrictions obtained
   in GR are directly extended to STT and HOG.

   I will not discuss the question of which conformal frame (Jordan,
   Einstein or some other) in the alternative theories should be regarded as
   a physical one, refering to our paper \cite{bm01} and references therein.

   The paper is organized as follows. \sect 2 gives the field equations.
   \sect 3 contains a brief description of purely vacuum structures in $D$
   dimensions with a cosmological constant. \sect 4 represents the results
   of Refs.\,\cite{vac1,vac2} on scalar vacuum in GR. Some no-go theorems
   are mentioned without proofs, but the main theorem on the possible
   horizons dispositions is given a new proof. Two examples of
   configurations admitted by the no-go theorems, are mentioned: a \bh\ with
   a nontrivial scalar field and a particlelike solution, both with
   non-positive-definite potentials. In \sect 5, the familiar STT $\mapsto$
   GR and HOG $\mapsto$ GR mappings are recalled and discussed, while in
   \sect 6 possible conformal continuations are studied.

   To conclude, with all theorems and examples at hand, we now have, even
   without solving the field equations, rather a clear picture of
   what can and what cannot be expected from \ssph\ scalar-vacuum
   configurations in various theories of gravity with various scalar field
   potentials.

   Throughout the paper all statements apply to \ssph\ configurations,
   and all relevant functions are assumed to be sufficiently smooth, unless
   otherwise indicated.

\section{Field equations}                           

    The field equations due to (\ref{act-d}) are
\bear
    \nabla^\alpha \nabla_\alpha \varphi + V_\varphi \eql 0,     \label{SE}
\\
    R\mN -\half \delta\mN\, R + T\mN \eql 0,                    \label{EE}
\ear
    where $V_\varphi \equiv dV/d\varphi$, $R\mN$ is the Ricci tensor and
    $T\mN$ is the energy-momentum tensor of the $\varphi$ field:
\beq
    T\mN = \varphi_{,\mu}\varphi^{,\nu}                         \label{EMT}
                - \half \delta\mN (\d\varphi)^2 + \delta\mN V (\varphi).
\eeq

    Consider a \ssph\ configuration, with the space-time structure
\beq
    \M^{\od+2} = \R_t \times \R_\rho \times \S^{\od},       \label{stru}
\eeq
    where $\R_t$ is the time axis, $\R_\rho\subset \R$ is the range
    of the radial coordinate $\rho$ and $\S^{\od}$ ($\od=D-2$) is
    a $\od$-dimensional sphere. The metric can be written in the form
\beq                                                          \label{ds-d}
    ds^2 = A(\rho) dt^2 - \frac{d\rho^2}{A(\rho)} - r^2(\rho)
				d\Omega_\od{}^2,
\eeq
    where $d\Omega_\od{}^2$ is the linear element on $\S^{\od}$
    of unit radius, and $\varphi=\varphi(\rho)$. (Without loss of
    generality, we suppose that large $\rho$ corresponds to large $r$.)
    Accordingly, \eq (\ref{SE}) and certain combinations of \eqs (\ref{EE})
    lead to
\bear
       (Ar^\od \varphi')' \eql r^\od V_\varphi;               \label{phi-d}
\\
       	      (A'r^\od)' \eql - (4/\od)r^\od V;               \label{00d}
\\
     	      \od r''/r  \eql -{\varphi'}^2;                 \label{01d}
\\
    A (r^2)'' - r^2 A'' + (\od-2)r'(2Ar' \!\al-\al\! A'r)
\nn
     		\eql 2(\od-1); 				    \label{02d}
\\
   \od(\od-1)(1-A{r'}^2) - \od A'rr' \eql -Ar^2{\varphi'}^2 + 2r^2 V,
\nnn
							      \label{int-d}
\ear
    where the prime denotes $d/d\rho$. Only three of these five equations
    are independent: the scalar equation (\ref{phi-d}) follows from
    the Einstein equations, while \eq (\ref{int-d}) is a first integral of
    the others. Given a potential $V(\varphi)$, this is a determined set of
    equations for the unknowns $r,\ A,\ \varphi$.

    The choice of the radial coordinate $\rho$ such that
    $g_{tt}g_{\rho\rho} = -1$ is convenient for a number of reasons. First,
    we are going to deal with horizons, which correspond to zeros of the
    function $A(\rho)$. One can notice that such zeros are regular
    points of \eqs (\ref{phi-d})--(\ref{int-d}), therefore one can jointly
    consider regions at both sides of a horizon. Second, in a close
    neighbourhood of a horizon $\rho$ varies (up to a positive constant
    factor) like manifestly well-behaved Kruskal-like coordinates used for
    an analytic continuation of the metric \cite{cold}. Third, with the same
    coordinate, horizons also correspond to regular points in geodesic
    equations \cite{cold}. Last but not least, this
    choice well simplifies the equations, in particular, (\ref{02d}) can be
    integrated, giving, for $\od \geq 2$,
\beq
    B'\equiv \biggl(\frac{A}{r^2}\biggr)' =
	- \frac{2(\od-1)}{r^{\od+2}} \int r^{\od-2} d\rho.      \label{A'd}
\eeq

    Our interest will be in the generic global behaviour of the
    solutions and the existence of BHs and globally regular configurations.

    In these issues, a crucial role belongs to {\it Killing horizons,\/}
    regular surfaces where the Killing vector $\d_t$ is null. For the
    metric (\ref{ds-d}), a horizon $\rho=h$ is a sphere of nonzero
    radius $r=r_h$ where $A=0$. The space-time regularity implies the
    finiteness of $T\mN$, so that $V$ and $A {\varphi'}^2$ are finite at
    $\rho=h$. The $C^2$-smoothness requirement for $r(\rho)$ at $\rho=h$
    means that $r''$ is finite, and (\ref{01d}) leads to $|\varphi'|
    <\infty$.

    The horizon is {\it simple\/} or {\it multiple\/} (or {\it
    higher-order\/}) according to whether the zero of the function $A(\rho)$
    is simple or multiple.  Thus, the Schwarzschild horizon is simple while
    the extreme Reissner-Nordstr\"om one is double.

    As usual, we shall call the space-time regions where $A>0$ and $A<0$
    {\it static\/} (R) and {\it nonstatic\/} (T) regions, respectively.
    The T regions represent homogeneous cosmological models of
    Kantowski-Sachs type. A simple or odd-order horizon separates a static
    region from a nonstatic one, whereas an even-order horizon separates two
    regions of the same nature. On the construction of Carter-Penrose
    diagrams, characterizing the causal structure of arbitrary static
    2-dimensional space-times [such as the ($t,\rho$) section of
    (\ref{ds-d})] see Refs.\,\cite{walker,br79} and more recent and more
    comprehensive papers \cite{katan,strobl}.

\section{GR: vacuum with a cosmological constant} 

\Picture{87}
{\unitlength=0.5mm
\special{em:linewidth 0.4pt}
\linethickness{0.4pt}
\begin{picture}(155.00,160.00)
(5,-4)
\put(15.00,71.00){\vector(1,0){142.00}}
\put(15.00,3.00){\vector(0,1){160.00}}
\bezier{568}(15.00,111.00)(104.00,114.00)(137.00,155.00)
\bezier{656}(29.00,159.00)(68.00,98.00)(136.00,160.00)
\bezier{568}(25.00,155.00)(46.00,112.00)(140.00,113.00)
\bezier{324}(138.00,148.00)(111.00,121.00)(71.00,106.00)
\bezier{484}(71.00,106.00)(24.00,87.00)(19.00,17.00)
\bezier{840}(145.00,109.00)(33.00,108.00)(21.00,11.00)
\bezier{1268}(24.00,9.00)(69.00,157.00)(137.00,10.00)
\bezier{1096}(30.00,8.00)(65.00,134.00)(133.00,8.00)
\bezier{924}(37.00,6.00)(64.00,111.00)(128.00,7.00)
\put(155.00,111.00){\makebox(0,0)[cc]{1a}}
\put(135.00,115.00){\makebox(0,0)[cb]{1b}}
\put(130.00,157.00){\makebox(0,0)[rb]{2b}}
\put(139.00,155.00){\makebox(0,0)[lc]{2a}}
\put(140.00,145.00){\makebox(0,0)[cc]{3}}
\put(140.00,106.00){\makebox(0,0)[ct]{4}}
\put(145.00,24.00){\makebox(0,0)[lb]{5b}}
\put(143.00,8.00){\makebox(0,0)[lb]{5a}}
\put(137.00,5.00){\makebox(0,0)[lc]{6a}}
\put(132.00,1.00){\makebox(0,0)[cb]{6b}}
\put(124.00,8.00){\makebox(0,0)[rb]{6c}}
\bezier{316}(81.00,105.00)(30.00,126.00)(20.00,148.00)
\bezier{460}(81.00,105.00)(123.00,86.00)(144.00,20.00)
\bezier{788}(15.00,111.00)(109.00,111.00)(141.00,13.00)
\put(151.00,111.00){\line(-1,0){135.91}}
\put(11.00,111.00){\makebox(0,0)[cc]{1}}
\end{picture}
}
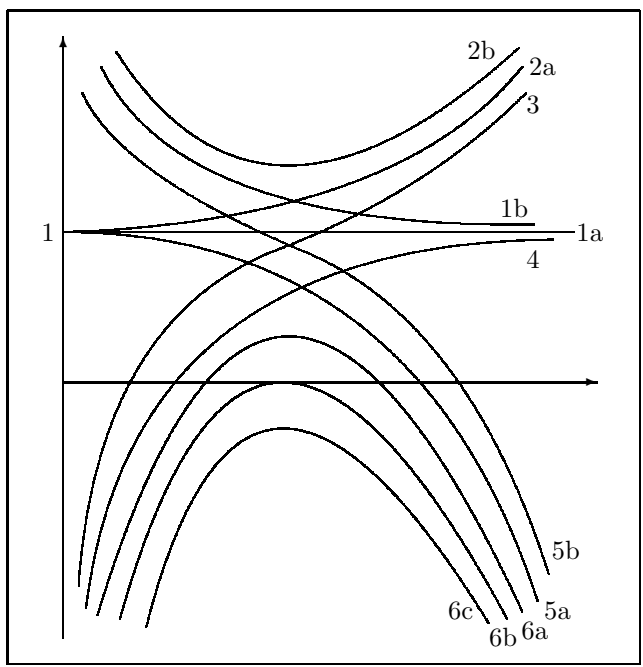
{The behaviour of $A(r)$, \eq(\ref{A-SdS}), for different values of $m$ and
	$\Lambda$.}

\Picture{100}
{\unitlength=0.75mm
\special{em:linewidth 0.4pt}
\linethickness{0.4pt}
\begin{picture}(130.00,136.00)
(23,8)
\put(35.00,134.00){\line(1,-1){10.00}}
\put(45.00,124.00){\line(-1,-1){10.00}}
\put(35.00,114.00){\line(0,1){21.00}}
\put(35.00,135.00){\line(1,-1){11.00}}
\put(46.00,124.00){\line(-1,-1){11.00}}
\put(35.00,113.00){\line(0,1){1.00}}
\bezier{120}(65.00,135.00)(55.00,124.00)(65.00,113.00)
\bezier{120}(65.00,135.00)(75.00,124.00)(65.00,113.00)
\put(95.00,135.00){\line(0,-1){21.00}}
\put(95.00,114.00){\line(1,0){20.00}}
\put(115.00,114.00){\line(0,1){21.00}}
\put(115.00,135.00){\line(-1,0){20.00}}
\put(95.00,135.00){\line(1,-1){20.00}}
\put(115.00,115.00){\line(1,-1){1.00}}
\put(116.00,114.00){\line(0,1){21.00}}
\put(116.00,135.00){\line(-1,0){1.00}}
\put(115.00,135.00){\line(-1,-1){20.00}}
\put(95.00,115.00){\line(-1,-1){1.00}}
\put(94.00,114.00){\line(0,1){21.00}}
\put(94.00,135.00){\line(1,0){1.00}}
\put(30.00,82.00){\line(1,1){10.00}}
\put(40.00,92.00){\line(1,0){20.00}}
\put(60.00,92.00){\line(1,-1){10.00}}
\put(70.00,82.00){\line(-1,-1){10.00}}
\put(60.00,72.00){\line(-1,0){20.00}}
\put(40.00,72.00){\line(-1,1){10.00}}
\put(39.00,72.00){\line(-1,1){10.00}}
\put(29.00,82.00){\line(1,1){10.00}}
\put(61.00,92.00){\line(1,-1){10.00}}
\put(71.00,82.00){\line(-1,-1){10.00}}
\put(40.00,72.00){\line(1,1){20.00}}
\put(40.00,92.00){\line(1,-1){20.00}}
\put(95.00,92.00){\line(1,0){20.00}}
\put(115.00,92.00){\line(0,-1){20.00}}
\put(115.00,72.00){\line(-1,0){20.00}}
\put(95.00,72.00){\line(0,1){20.00}}
\put(95.00,92.00){\line(1,-1){20.00}}
\put(115.00,72.00){\line(0,-1){1.00}}
\put(115.00,71.00){\line(-1,0){20.00}}
\put(95.00,71.00){\line(0,1){1.00}}
\put(95.00,72.00){\line(1,1){20.00}}
\put(115.00,92.00){\line(0,1){1.00}}
\put(115.00,93.00){\line(-1,0){20.00}}
\put(95.00,93.00){\line(0,-1){1.00}}
\bezier{136}(65.00,136.00)(77.00,124.00)(65.00,112.00)
\put(23.00,48.00){\line(1,0){49.00}}
\put(22.00,28.00){\line(1,0){52.00}}
\put(71.00,33.00){\line(-1,-1){5.00}}
\put(66.00,28.00){\line(-1,1){20.00}}
\put(46.00,48.00){\line(-1,-1){20.00}}
\put(26.00,28.00){\line(-1,1){4.00}}
\put(22.00,44.00){\line(1,1){4.00}}
\put(26.00,48.00){\line(1,-1){20.00}}
\put(46.00,28.00){\line(1,1){20.00}}
\put(66.00,48.00){\line(1,-1){4.00}}
\put(46.00,49.00){\line(1,0){20.00}}
\put(26.00,49.00){\line(-1,0){4.00}}
\put(22.00,27.00){\line(1,0){4.00}}
\put(46.00,27.00){\line(1,0){20.00}}
\put(87.00,50.00){\line(1,0){43.00}}
\put(130.00,40.00){\line(-1,0){42.00}}
\put(88.00,44.00){\line(1,-1){4.00}}
\put(92.00,40.00){\line(1,1){10.00}}
\put(102.00,50.00){\line(1,-1){10.00}}
\put(112.00,40.00){\line(1,1){10.00}}
\put(122.00,50.00){\line(1,-1){8.00}}
\put(130.00,51.00){\line(-1,0){43.00}}
\bezier{104}(98.00,21.00)(109.00,29.00)(119.00,21.00)
\bezier{120}(97.00,21.00)(109.00,31.00)(120.00,21.00)
\bezier{96}(98.00,21.00)(109.00,15.00)(119.00,21.00)
\put(38.00,107.00){\makebox(0,0)[cc]{1a, 1b}}
\put(65.00,107.00){\makebox(0,0)[cc]{2a, 2b}}
\put(105.00,107.00){\makebox(0,0)[cc]{3}}
\put(50.00,64.00){\makebox(0,0)[cc]{4}}
\put(105.00,64.00){\makebox(0,0)[cc]{5a, 5b}}
\put(45.00,18.00){\makebox(0,0)[cc]{6a}}
\put(98.00,34.00){\makebox(0,0)[cc]{6b}}
\put(98.00,14.00){\makebox(0,0)[cc]{6c}}
\put(39.00,124.00){\makebox(0,0)[cc]{R}}
\put(65.00,124.00){\makebox(0,0)[cc]{R}}
\put(99.00,125.00){\makebox(0,0)[cc]{R}}
\put(111.00,125.00){\makebox(0,0)[cc]{R}}
\put(40.00,82.00){\makebox(0,0)[cc]{R}}
\put(60.00,82.00){\makebox(0,0)[cc]{R}}
\put(99.00,82.00){\makebox(0,0)[cc]{R}}
\put(112.00,82.00){\makebox(0,0)[cc]{R}}
\put(26.00,38.00){\makebox(0,0)[cc]{R}}
\put(46.00,38.00){\makebox(0,0)[cc]{R}}
\put(66.00,38.00){\makebox(0,0)[cc]{R}}
\put(50.00,88.00){\makebox(0,0)[cc]{T${}_+$}}
\put(105.00,118.00){\makebox(0,0)[cc]{T${}_+$}}
\put(105.00,88.00){\makebox(0,0)[cc]{T${}_+$}}
\put(36.00,32.00){\makebox(0,0)[cc]{T${}_+$}}
\put(56.00,44.00){\makebox(0,0)[cc]{T${}_+$}}
\put(92.00,46.00){\makebox(0,0)[cc]{T${}_+$}}
\put(112.00,46.00){\makebox(0,0)[cc]{T${}_+$}}
\put(102.00,43.00){\makebox(0,0)[cc]{T${}_+$}}
\put(122.00,43.00){\makebox(0,0)[cc]{T${}_+$}}
\put(50.00,75.00){\makebox(0,0)[cc]{T${}_-$}}
\put(105.00,75.00){\makebox(0,0)[cc]{T${}_-$}}
\put(105.00,131.00){\makebox(0,0)[cc]{T${}_-$}}
\put(36.00,44.00){\makebox(0,0)[cc]{T${}_-$}}
\put(56.00,31.00){\makebox(0,0)[cc]{T${}_-$}}
\put(109.00,21.00){\makebox(0,0)[cc]{T${}_+$}}
\end{picture}
}
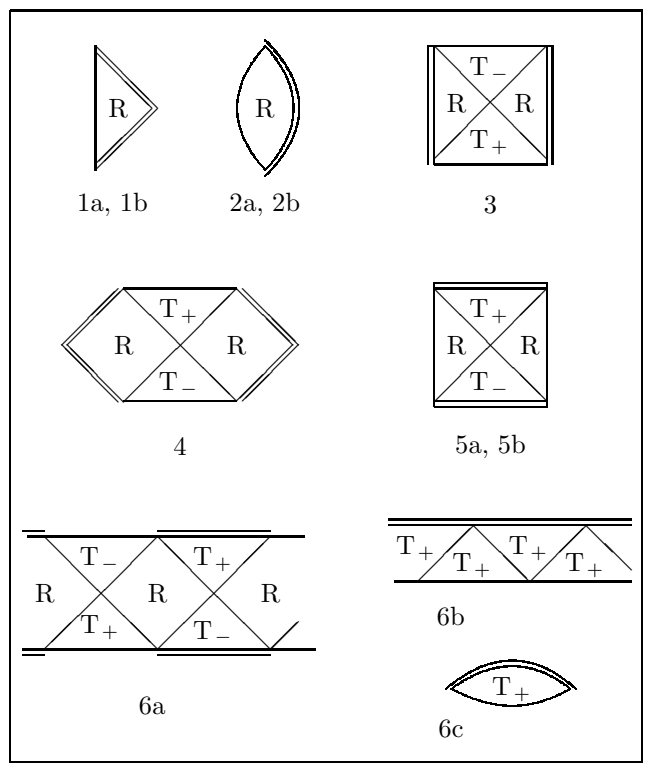
{Carter-Penrose diagrams for different cases of the metric (\ref{g-SdS}),
	(\ref{A-SdS}), labelled according to \fig 1. The R and T
	letters correspond to R and T space-time regions; T${}_+$ and
	T${}_-$ denote expanding and contracting T region (i.e., with $r$
	increasing and decreasing with time, respectively). Single
	lines on the border of the diagrams denote $r=0$, double lines ---
	$r=\infty$. Diagrams 6b and 6c are drawn for the case of expanding
	KS cosmologies; to obtain diagrams for contracting models, one
	should merely interchange $r=0$ and $r=\infty$ and replace T${}_+$
	with T${}_-$.}

    In case $\varphi=\const,\ V = \const = \Lambda$, one can without loss
    of generality take $r=\rho$, then \eq (\ref{int-d}) becomes a linear
    first-order equation with respect to $A(r)$ whose integration gives
\beq                                                          \label{A-SdS}
     A(r) = 1 - \frac{2m}{r^{\od-1}} - \frac{2\Lambda r^2}{\od(\od+1)}.
\eeq
    The metric has the form
\beq                                                          \label{g-SdS}
    ds^2 = A(r) dt^2 - \frac{dr^2}{A(r)} - r^2\,d\Omega_\od{}^2.
\eeq
    This is the multidimensional \Sch-\dS\ solution.
    Its special cases correspond to the \Sch\ ($\od=2$, $\Lambda=0$) and
    Tangherlini ($\od$ arbitrary, $\Lambda=0$) solutions and the \dS\
    solution in arbitrary dimension ($m=0$). The latter is often called
    anti-\dS\ in case $\Lambda <0$.

    The different qualitative behaviours of $A(r)$ for different values of
    $\Lambda$ and $m$ correspond to the following structures:
\begin{enumerate}\itemsep -1.5pt
\item
  $\Lambda = 0,\ m\leq 0$: curves 1a and 1b in \fig 1, diagram 1 in \fig 2
  (Minkowski and $m<0$ \Sch, respectively).
\item
  $\Lambda < 0,\ m\leq 0$: curves 2a and 2b in \fig 1, diagram 2 in \fig 2
  (AdS and $m<0$ \Sch-AdS).
\item
  $\Lambda < 0,\ m > 0$: curve 3 in \fig 1, diagram 3 in \fig 2
  (\Sch-AdS).
\item
  $\Lambda = 0,\ m > 0$: curve 4 in \fig 1, diagram 4 in \fig 2 (\Sch).
\item
  $\Lambda > 0,\ m\leq 0$: curves 5a and 5b in \fig 1, diagram 5 in \fig 2
  ($m<0$ \Sch-\dS).
\item
  $\Lambda > 0,\ m > 0$: curves 6a, 6b and 6c in \fig 1, and the
  corresponding diagrams in \fig 2 (\Sch-\dS\ in case 6a and Kantowski-Sachs
  cosmologies in cases 6b and 6c).
\end{enumerate}
    The centre $r=0$ is regular for $m=0$ and singular for $m\ne 0$.

    In case 6, given a particular value of $\Lambda > 0$, the solution
    behaviour depends on the mass parameter $m$. When $m$ is smaller than
    the critical value
\beq
 	m_{\rm cr} = \frac{1}{\od+1}
		\biggl[ \frac{\od(\od-1)}{2\Lambda}\biggr]^{(\od-1)/2},
\eeq
    there are two horizons, the smaller one being interpreted as a \bh\
    horizon and the greater one as a cosmological horizon. If $m=m_{\rm
    cr}$, these two horizons merge, and one has two homogeneous T regions
    separated by a double horizon. Lastly, the solution with
    $m > m_{\rm cr}$ is purely cosmological, having no Killing horizons.

\section{Scalar vacuum in GR. No-go theorems and global structures} 

\subsection{Regular models without a centre?}  

    The first important restriction for the system (\ref{act-d})
    in the general case is that such configurations as wormholes, horns or
    flux tubes do not exist under our assumptions.

    For the metric (\ref{ds-d}), a (traversable, Lorentzian) {\it wormhole\/}
    is, by definition, a configuration with two asymptotics at which $r\to
    \infty$, hence with $r(\rho)$ having at least one regular minimum. A
    {\it horn\/} is a region where, as $\rho$ tends to some value $\rho^*$,
    $r(\rho)\ne \const$ and $g_{tt} = A$ have finite limits while the length
    integral $l = \int d\rho/A $ diverges. In other words, a horn is an
    infinitely long $(\od+1)$-dimensional ``tube'' of finite radius, with the
    clock rate remaining finite everywhere. Such ``horned particles'' were,
    in particular, discussed as possible remnants of black hole evaporation
    \cite{banks}.  Lastly, a {\it flux tube\/} is a configuration with $r =
    \const$.

\Theorem{Theorem 1}
    {The field equations due to (\ref{act-d}) do not admit
    (i) solutions where the function $r(\rho)$ has a regular minimum,
    (ii) solutions describing a horn, and
    (iii) flux-tube solutions with $\varphi\ne\const$.
    }

    The formulation of the theorem and its proof \cite{vac1, vac2} do
    not refer to any kind of asymptotic, therefore wormhole throats or horns
    are absent in solutions having any large $r$ behaviour --- flat, de
    Sitter or any other, or having no large $r$ asymptotic at all.

    It also follows that the full range of the $\rho$ coordinate covers
    all values of $r$, from the centre ($\rho=\rho_c$, $r(\rho_c)=0$),
    regular or singular, to infinity, unless (which is not excluded) there
    is a singularity at finite $r$ due to a ``pathological'' choice of
    the potential.

    The latter opportunity deserves attention since, being singular at zero
    or finite $r$, the space-time may in principle be still geodesically
    complete. In other words, any geodesic can only reach the singularity
    at an infinite value of its canonical parameter. No freely
    moving particle can then attain such a singularity (to be called a {\it
    remote singularity\/} in finite proper time.
    Examples of remote singularities are known in solutions of 2-dimensional
    gravity \cite{zasl}).

    We can, however, state the following \cite{vac2}:

\Theorem{Theorem 2}
    {If a solution to \eqs (\ref{phi-d})--(\ref{int-d}) has a spatial
    asymptotic ($r\to\infty$), it cannot contain a remote singularity at
    $r < \infty$. }

    Thus remote singularities can only exist in configurations like closed
    cosmological models, unable to describe isolated bodies observable from
    outside.

\subsection{Global structures}                       

    Now, taking into account Theorem 1, the global space-time structure
    corresponding to any particular solution is unambiguously determined
    (up to identification of isometric surfaces, if any) by the
    disposition of static ($A>0$) and nonstatic ($A < 0$) regions.
    The following theorem severely restricts the choice of horizon
    dispositions in the theory under study.

\Theorem{Theorem 3}
   {Consider solutions of the theory (\ref{act-d}), $D\geq 4$, with the
    metric (\ref{ds-d}) and $\varphi=\varphi(\rho)$. Let there be a static
    region $a < \rho < b \leq \infty$. Then:
\begin{description}\itemsep -2pt
\item [(i)]
    all horizons are simple;
\item [(ii)]
    no horizons exist at $\rho < a$ and at $\rho > b$.
\end{description} }

\noi{\bf Proof.}
    \eq (\ref{02d}) may be rewritten as follows:
\beq
	r^4 B'' + (\od+2) r^3 r'B' = -2(\od-1)                 \label{eq-B}
\eeq
    where $B(\rho) = A/r^2$. Evidently, zeros of $A(\rho)$ such that $r\ne
    0$ are zeros of the same order of the function $B(\rho)$. By
    (\ref{eq-B}), $B(\rho)$ cannot have a regular minimum since
    $B'=0$ implies $B''(h) = -2(\od-1)/r^4 <0$.

    Therefore, if $\rho=h$ is a horizon of at least order,
    $B(h)=B'(h)=0$, it is a maximum of $B(\rho)$, hence a double
    horizon separating two T regions. The absence of regular minima of $B$
    then means that $B<0$ for all $\rho\ne h$, i.e., there is no static
    region --- item (i) is proved.

    Consider now the boundary $\rho=a$ of the static region. If $r(a)=0$,
    it is the centre; be it regular or singular, it is then the left
    boundary of the range of $\rho$. If $r(a)\ne 0$, then it is a
    simple horizon: $B(a)=0$, $B'(a) > 0$. Since $B$ has no minima, it is
    negative and non-decreasing for all $\rho <a$, i.e., there is no
    horizon. In a similar way one obtains that horizons are absent to the
    right of $b$, thus completing the proof.

\medskip
    This theorem shows that the possible disposition of zeros of the
    function $A(\rho)$ (or $B(\rho)$) is the same as in the vacuum case
    described in \sect 3. Therefore the list of possible global structures
    is also the same.

    Theorem 3 shows, in particular, that the attractive idea of replacing
    the black hole singularity by a nonsingular vacuum core [16--20]
    cannot be realized in the theory (\ref{act-d}). Indeed, such a BH, with
    any large $r$ behavior, must have static regions at small and large $r$,
    separated by at least two simple or one double horizon, i.e., the
    function $B(\rho)$ must have at least one regular minimum. This is
    impossible due to \eq (\ref{eq-B}).

    More generally, one can conclude that if spatial infinity is static,
    there is at most one simple horizon; the same is true if the centre is
    in a static region.

\subsection*{Special case: (2+1)-dimensional gravity}

    In 3 dimensions we have $\od=1$, and integration of (\ref{02d}) leads to
    an expression simpler than (\ref{A'd}):
\beq
	B'\equiv (A/r^2)' = C/r^3, \cm C= \const.    		\label{A'3}
\eeq

    In Theorem 1, items (i) and (iii) hold due to \eq (\ref{01d}), as
    before. Still, the proof of item (ii) does not work: a horn is
    possible if, in (\ref{A'3}), $C=0$. Though, due to $r''<0$, the horn
    radius $r^*$ is the maximum of $r(\rho)$, so that a horned configuration
    has no large $r$ asymptotic.

    By virtue of (\ref{A'3}), $B'$ has a constant sign coinciding with
    $\sign C$, and, instead of Theorem 3, we have a still more severe
    restriction:

\Theorem {Theorem 3a}
    {A static, circularly symmetric configuration in the theory
    (\ref{act-d}), $D=3$, has either no horizon or one simple horizon.}

    Accordingly, the list of possible global structures is even shorter
    than the previous one: the structures corresponding to the curves 6a and
    6b are absent.

\subsection{4-dimensional GR: restrictions and examples}   

    The above theorems did not use any assumptions on the asymptotic
    behaviour of the solutions or the shape and even sign of the potential.
    Let us now mention some more specific but also very significant results
    for positive-semidefinite potentials.

    Consider, for simplicity, $D=4$. The field functions at a regular centre
    and at a flat asymptotic (if they exist) behave as follows.

    A {\bf regular centre}, where $r=0$, implies a finite time rate and
    local spatial flatness. This means that at some finite $\rho=\rho_c$
\bear
    	 A{r'}^2 \to 1,   \cm 	  A = A_c + O (r^2),           \label{c1}
\ear
    where $A_c = A(\rho_c)$ and $r'(\rho_c)$ are finite and positive.
    Moreover, the values of $V$, $\varphi$ and $\varphi'$ should be finite
    there. Then from (\ref{01d}) and (\ref{A'd}) one obtains:
\beq
    	  r''(\rho_c) = 0; \cm    \rho_0 = \rho_c.             \label{c2}
\eeq

    At a {\bf flat asymptotic}, the metric should behave as the
    Schwarzschild one with a certain mass $M$, while $\varphi$ should tend
    to a finite value. Thus we have
\bearr
    \rho\to\infty; \qquad r'\to 1; \qquad
	  		A(\rho) = 1-\frac{2M}{\rho} + O(\rho^{-2});
\nnn
    \varphi' = o (\rho^{-3/2});   \qquad    V = o(\rho^{-3});   \label{a1}
\ear

    One of the known restrictions is the no-hair theorem:

\Theorem{Theorem 4 (no-hair)}
    {Suppose $V \geq 0$. Then the only asymptotically flat BH solution to
    \eqs (\ref{phi-d})--(\ref{int-d}) in the range $(h,\infty)$ (where
    $\rho=h$ is the event horizon) comprises the Schwarzschild metric,
    $\varphi =\const$ and $V\equiv 0$.  }

    This theorem was first proved by Bekenstein \cite{bek72} for the case of
    $V(\varphi)$ without local maxima and was later refined for any
    $V \geq 0$ and for certain more general Lagrangians --- see e.g.
    \Ref{bek98} for proofs and references.

    Another restriction can be called the generalized Rosen
    theorem (G. Rosen \cite{rosen} studied similar restrictions for
    flat-space nonlinear field configurations):

\Theorem{Theorem 5 \cite{brsh}}
    {An asymptotically flat solution with positive mass $M$ and a regular
    centre is impossible if $V(\varphi)\geq 0$.          }

    The above theorems leave some opportunities of interest, in particular:
\begin{enumerate} \itemsep -1.5pt
\item
    BHs with $\varphi\ne \const$, potentials $V(\varphi)\geq 0$ but
    non-flat large $r$ asymptotics;
\item
    asymptotically flat BHs with $\varphi\ne \const$ but at least
    partly negative potentials $V(\varphi)$;
\item
    asymptotically flat particlelike solutions (solitons) with positive mass
    but at least partly negative potentials $V(\varphi)$.
\end{enumerate}
    That such solutions do exist, one can prove by presenting proper
    examples. For item 1, such examples have been given in \Ref{Mann95},
    where, among other results, BHs with non-flat asymptotics were found for
    the Liouville ($V=2\Lambda\e^{2b\varphi}$) and double Liouville
    ($V=2\Lambda_1\e^{2b_1\varphi} + 2\Lambda_2\e^{2b_2\varphi}$)
    potentials, where the $\Lambda$'s and $b$'s are positive constants.

    Special analytical solutions to \eqs (\ref{phi-d})--(\ref{int-d}) for
    $\od=2$, exemplifying items 2 and 3 (Appendix B), were given in
    \Ref{vac2}. Unlike \Ref{Mann95}, where special solutions were sought for
    by making the ansatz $r(\rho)\propto \rho^N$, $N=\const$ (in our
    notation), we have used in \Ref{vac2} the following approach.  Suppose
    $V(\varphi)$ is one of the unknowns. Then our set of equations is
    underdeterminate, and we can choose one of the unknowns arbitrarily
    trying to provide the proper behaviour of the solution. Thus, one can
    choose a particular function $r(\rho)$:  assigning it arbitrarily and
    substituting into (\ref{A'd}), by single integration we obtain
    $A(\rho)$, after which $\varphi(\rho)$ and $V(\rho)$ are determined from
    (\ref{01d}) and (\ref{00d}), respectively.  Thus $V(\varphi)$ is
    obtained in a parametric form; it can be made explicit if
    $\varphi(\rho)$ resolves with respect to $\rho$.

    A \bh\ solution was obtained \cite{vac2} by choosing
\beq
	r(\rho) = \sqrt{\rho^2 -a^2}, \cm a =\const >0,
\eeq
    whereas a solitonic solution with positive mass was found under the
    assumption
\beq
	r^2(\rho) = \frac{a^2}{\rho^2} \frac{\tanh (a/\rho + c)}{\tanh c},
\eeq
    with some positive constants $a$ and $c$. These assumptions lead
    to potentials $V(\varphi) <0$ which do not seem quite realistic.
    However, the purpose of giving these examples was to merely demonstrate
    the existence of such kinds of solutions. After this demonstration,
    it makes sense to seek similar solutions for more plausible potentials
    using numerical methods.

\subsection{More general Lagrangians in GR. Sigma models}        

    One can notice that Theorems 1--3 actually rest on two Einstein
    equations, (\ref{01d}) or (\ref{02d}), which in turn follow from
    the properties of the energy-momentum tensor. Namely, the property
    $T^t_t - T^\rho_\rho \geq 0$ expresses the validity of
    the null energy condition for systems with the metric (\ref{ds-d}).
    The corresponding Einstein equation then implies $r'' \leq 0$.
    \eq (\ref{02d}), which leads to Theorem 3, follows from the property
\beq
        T^t_t = T^\theta_\theta                              \label{T02}
\eeq
    where $\theta$ is any of the coordinate angles that parametrize the
    sphere $\S^{\od}$.

    Therefore these three theorems hold for all kinds of matter whose
    energy-momentum tensors satisfy these two conditions.

    Consider, for instance, the following action, more general than
    (\ref{act-d}):
\beq                                                          \label{act'}
	S = \int d^D x \,\sqrt{-g} [R + F(I, \varphi)]
\eeq
    where $I = (\d\varphi)^2$ and $F(I, \varphi)$ is an arbitrary function.
    The scalar field energy-momentum tensor is
\beq
	T\mN = \frac{\d F}{\d I}
         	\varphi_{,\mu}\varphi^{,\nu}                  \label{EMT'}
        	               + \half \delta\mN F (\varphi).
\eeq
    In the \ssph\ case, \eq (\ref{T02}) holds automatically due to
    $\varphi=\varphi(\rho)$, while the null energy condition holds
    as long as $\d F/\d I \geq 0$, which actually means that the kinetic
    energy is nonnegative. Under this condition, all Theorems 1--3 are
    valid for the theory (\ref{act'}). Otherwise Theorem 3 alone holds; it
    correctly describes the $\rho$ dependence of $A$ and consequently
    the possible horizons disposition, but the situation is more complex
    due to possible non-monotonicity of $r(\rho)$.

    Another important and frequently discussed class of theories are
    the so-called sigma models, where a set of $N$ scalar fields
    $\varphi = \{\varphi^a\}$, $a=\overline{1,N}$ are considered as
    coordinates of a target space with a certain metric $G_{ab}=
    G_{ab}(\varphi)$. The scalar vacuum action is then written in the form
\beq \nq
    S_\sigma = \int d^D x \sqrt{|g|}[R +
     		G_{ab}g\MN \d_\mu\varphi^a \d_\nu\varphi^b -2V(\varphi)]
\eeq
    where, in general, $G_{ab}$ and $V$ are arbitrary functions of
    $N$ variables, but in practice they possess symmetries that follow from
    the nature of specific systems.

    It is easily seen that, provided the metric $G_{ab}(\varphi)$
    is positive-definite, Theorems 1--3 for \ssph\ configurations are
    valid as before.

    If $G_{ab}$ is not positive-definite, or if some of $\varphi^a$ are
    allowed to be imaginary, only Theorem 3 holds.

\section{Scalar-tensor and higher-order gravity}              

    Other extensions of the above results concern theories connected with
    (\ref{act-d}) and (\ref{act'}) via $\varphi$-dependent conformal
    transformations, such as scalar-tensor theories (STT) and the
    so-called high-order gravity (HOG) (e.g., with the Lagrangian function
    $f(R)$).

    Above all, it should be noted that if a space-time $\M[g]$ with the
    metric (\ref{ds-d}) is conformally mapped into another space-time
    $\oM[\og]$, equipped with the same coordinates, according to the law
\beq
    g\mn = F(\rho) \og\mn,                                   \label{g-conf}
\eeq
    then it is easily verified that a horizon $\rho=h$ in $\M$ passes into a
    horizon of the same order in $\oM$, (ii) a centre ($r=0$), an asymptotic
    ($r\to \infty$) and a remote singularity in $\M$ passes into a center,
    an asymptotic and a remote singularity, respectively, in $\oM$ if the
    conformal factor $F(\rho)$ is regular (i.e., finite, at least
    C${}^2$-smooth and positive) at the corresponding values of $\rho$.
    A regular centre passes to a regular centre and a flat asymptotic
    to a flat asymptotic under evident additional requirements, but we will
    not concentrate on them here.

    The general (Bergmann-Wagoner-Nordtvedt) STT action in $D$ dimensions
    can be written as follows:
\bearr
     S_{\rm STT} = \int d^D x \sqrt{|g|}                 \label{act-J}
		   [f(\phi) R
\nnn \inch
		   + h(\phi) (\d\phi)^2 -2U(\phi) + L_m],
\ear
    where $f$, $h$ and $U$ are functions of the scalar field $\phi$ and
    $L_m$ is the matter Lagrangian. The metric $g\mn$ here corresponds to
    the so-called Jordan conformal frame. The standard transition to the
    Einstein frame \cite{wagon},
\bear
    g\mn \eql F(\varphi) \og\mn,\cm  F=|f|^{-2/(D-2)},    \label{g-wag}
\\
    \frac{d\varphi}{d\phi} \eql \frac{\sqrt{|l(\phi}|}{f(\phi)},
\qquad                                                    \label{phi-wag}
    l(\phi) \eqdef fh + \frac{D-1}{D-2}\biggl(\frac{df}{d\phi}\biggr)^2,
\ear
    removes the nonminimal scalar-tensor coupling express\-ed in a
    $\phi$-dependent coefficient before $R$. Putting $L_m=0$ (vacuum), one
    can write the action (\ref{act-J}) in terms of the new metric $\og\mn$
    and the new scalar field $\varphi$ as follows (up to a boundary term):
\beq
     S_{\rm E} = \eta_f\int d^D x \sqrt{|\og|}               \label{act-E}
    		[R_{\rm E} + \eta_l(\d\varphi)^2 -2V(\varphi)],
\eeq
    where $R_{\rm E}$ and $(\d\varphi)^2$ are calculated using $\og\mn$,
\beq
       V(\varphi) = \eta_f F^2 (\varphi)\, U(\phi),           \label{V-E}
\eeq
    and $\eta_{l,f}$ are sign factors:
\beq
     \eta_l = \sign l(\phi), \cm \eta_f = \sign f(\phi).      \label{etas}
\eeq

    Note that $\eta_l = -1$ corresponds to the so-called anomalous STT,
    with a wrong sign of scalar field kinetic energy, while $\eta_f=-1$
    means that the effective gravitational constant in the Jordan frame is
    negative. So the normal choice of signs is $\eta_{l,f}=1$.

    The action (\ref{act-E}) obviously coincides with (\ref{act-d}) up to the
    factor $\eta_l$. Thus \eq (\ref{T02}) holds, and we can assert
    that, for \ssph\ configurations, Theorem 3 is valid for the
    Einstein-frame metric $\og\mn$.

    Theorems 1 and 2 hold for $\og\mn$ only in the ``normal'' case
    $\eta_l=1$; let us adopt this restriction.

    The validity of the theorems for the Jordan-frame metric $g\mn$ depends
    on the nature of the conformal mapping (\ref{g-wag}) between the
    space-times $\M [g]$ (Jordan) and $\oM [\og]$ (Einstein). There are
    four variants:

\def\map{\ \longleftrightarrow\ }
\begin{description}\itemsep -2pt
\item[I.]
	$\M \ \map\ \oM$,
\item[II.]
	$\M \ \map\ (\oM_1 \subset \oM$),
\item[III.]
	$(\M_1 \subset \M) \ \map\ \oM$,
\item[IV.]
	$(\M_1 \subset \M) \ \map\ (\oM_1 \subset \oM$),
\end{description}
    where $\map$ denotes a diffeomorphism preserving the metric signature.
    The last three variants are possible if the conformal factor $F$
    vanishes or blows up at some values of $\rho$, which then mark the
    boundary of $\M_1$ or $\oM_1$.

    Theorem 3 on horizon dispositions is obviously valid in $\oM$ in
    cases I and II. In case III or IV, the whole space-time $\oM$ or its
    part is put into correspondence to only a part $\M_1$ of $\M$, and,
    generally speaking, anything, including additional horizons, can appear
    in the remaining part $\M_2=\M \setminus \M_1$ of the Jordan-frame
    space-time. The existence of such a region $\M_2$ will be refered to as a
    {\sl conformal continuation\/} of $\oM$ in $\M$.

    Theorem 1 cannot be directly transferred to $\M$ in any case except the
    trivial one, $F=\const$. It is only possible to assert, without
    specifying $F(\varphi)$, that wormholes as global entities are impossible
    in $\M$ in cases I and II if the conformal factor $F$ is finite in the
    whole range of $\rho$, including the boundary values. Indeed, if we
    suppose that there is such a wormhole, it will immediately follow that
    there are two large $r$ asymptotics and a minimum of $r(\rho)$
    between them even in $\oM$, in contrast to Theorem 1 which is valid
    there.

    Theorem 2 also evidently holds in $\M$ in cases I and II if the
    conformal factor $F$ is regular in the whole range of $\rho$, including
    the boundary values.

    Another class of theories conformally equivalent to (\ref{act-d}) is
    the so-called higher-order gravity (HOG) with the vacuum action
\beq
     S_{\rm HOG} = \int d^D x \sqrt{|g|}f(R)               \label{act-hog}
\eeq
    where $f$ is a function of the scalar curvature $R$ calculated for the
    metric $g\mn$ of a space-time $\M$. In accord with the weak field limit
    $f\sim R$ ar small $R$, let us assume $f(R) >0$ and $f_R \eqdef df/dR
    >0$, at least in a certain range of $R$ including $R=0$.
    The conformal mapping $\M[g] \mapsto \oM[\og]$ with
\beq
    g\mn = F(\varphi) \og\mn,\cm  F= f_R^{-2/(D-2)}, \label{g-hog}
\eeq
    transforms the ``Jordan-frame'' action (\ref{act-hog}) into the
    Einstein-frame action (\ref{act-d}) where
\bear                                                      \label{phi-hog}
    \varphi\eql \sqrt{\frac{D-1}{D-2}}\log f_R,
\\
    2V(\varphi) \eql f_R^{-D/(D-2)} (Rf_R - f). 	   \label{V-hog}
\ear
    The field equations due to (\ref{act-hog}) after this substitution
    turn into the field equations due to (\ref{act-d}).

    All the above observations on the validity of Theorems 1--3
    in STT equally apply to higher-order gravity.


    In what follows, we will first consider an exactly soluble example with
    a conformally coupled scalar field in GR, when the mapping follows
    variant III, and the conformal continuation creates a horizon or a
    wormhole throat outside $M_1$. Then we will obtain necessary
    conditions for the occurence of conformal continuations in
    4-dimensional STT (\ref{act-J}) and HOG (\ref{act-hog}), showing that
    this phenomenon is only possible under special requirements to the
    particular choice of these theories.

\section {Conformal continuations}                              

\subsection {Conformal scalar field in GR: \bh{}s and wormholes} 

    Conformal scalar field in GR can be viewed as a special case of STT,
    such that, in \eq (\ref{act-J}), $D=4$ and
\beq
    f(\phi) = 1 - \phi^2/6, \qquad  h(\phi)=1, \qquad U(\phi) =0.
\eeq
    After the conformal mapping
\bear                                                       \label{g-con}
    g\mn \eql F(\varphi)\og\mn, \qquad
     				F(\varphi)=\cosh^2(\varphi/\sqrt{6}),
\\
     \phi \eql \sqrt{6} \tanh (\varphi/\sqrt{6}),           \label{phi-con}
\ear
    we obtain the action (\ref{act-d}) with $D=4$ and $V\equiv 0$. The
    latter describes a minimally coupled massless scalar field in GR, and
    the corresponding \ssph\ solution is well-known: it is the Fisher
    solution \cite{fisher}.  It is convenient to write it using the harmonic
    radial coordinate $u$ specified by the condition \cite{br73} $|g_{uu}| =
    g_{tt}g_{\theta\theta}^2$ ($u$ behaves as $1/r$ at large $r$):
\bear
    ds^2_{\rm E} \eql \e^{-2mu}dt^2
     		- \frac{k^2\e^{2mu}}{\sinh^2(ku)}
        \biggl[\frac{k^2 du^2}{\sinh^2(ku)} + d\Omega^2\biggr],
\nn
     \varphi \eql \sqrt{6}C (u+u_0),                       \label{fish}
\ear
    where the subscript ``E'' stands for the Einstein frame,
    $m$ (the mass), $C$ (the scalar charge), $k>0$ and $u_0$ are
    integration constants, and $k$ is expressed in terms of $m$ and $C$:
\beq
	k^2 = m^2 + 3C^2.                                 \label{k-fish}
\eeq
    The previously used coordinate $\rho$, corresponding to the metric
    (\ref{ds-d}), $D=4$, is $\rho = 2k/(1-\e^{-2ku})$,
    and the metric in terms of $\rho$ has the form
\bearr                                                   \label{ds-fish}
    ds^2_{\rm E}= (1 -2k/\rho)^{m/k} dt^2
\nnn \quad
    - (1-2k/\rho)^{-m/k}
          	  \bigl[d\rho^2 + \rho^2(1-2k/\rho)d\Omega^2\bigr].
\ear

    This solution is asymptotically flat at $u\to 0$ ($\rho\to\infty$), has
    no horizon when $C\ne 0$ (as should be the case according to the no-hair
    theorem) and is singular at the centre ($u\to \infty$, $\rho\to 2k$,
    $\varphi\to\infty$). It turns into the \Sch\ solution when $C=0$.

    The ``Jordan-frame'' solution is described by the metric
    $ds^2 = F(\varphi) ds_{\rm E}^2$ and the $\phi$ field according to
    (\ref{phi-con}). It is the conformal scalar field solution \cite{bbm70,
    bek74}, its properties are more diverse and can be presented as
    follows (putting, for definiteness, $m>0$ and $C>0$):

\medskip\noi
{\bf 1.} $C < m$. The metric behaves qualitatively as in the Fisher
    solution:  it is flat at $u\to 0$, and both $g_{tt}$ and
    $r^2=|g_{\theta\theta}|$ vanish at $u\to \infty$ --- a singular
    attracting centre. A difference is that here the scalar field is finite:
    $\phi\to \sqrt{6}$.

\medskip\noi
{\bf 2.} $C > m$. Instead of a singular centre, at $u\to\infty$ one has a
    repulsive singularity of infinite radius: $g_{tt}\to \infty$
    and $r^2 \to \infty$. Again $\phi\to \sqrt{6}$ as $u\to \infty$.

\medskip\noi
{\bf 3.} $C = m$. In this case the metric and $\phi$ are regular
    at $u = \infty$; a continuation across this regular sphere may be
    achieved using a new coordinate, e.g.,
\beq
	y=\tanh (mu).                                         \label{u-y}
\eeq
    The solution acquires the form
\bear
     ds^2 \eql \frac{(1+yy_0)^2}{1-y_0^2}\biggl[\frac{dt^2}{(1+y)^2}
\nnn \inch
               -\frac{m^2(1+y)^2}{y^4}(dy^2+y^2 d\Omega^2)\biggr],
\nn
     \phi \eql \sqrt{6} \frac{y+y_0}{1 + yy_0},          \label{con-y}
\ear
     where $y_0 = \tanh(m u_0)$. The range $u\in \R_+$, describing the
     whole manifold $\oM$ in the Fisher solution, corresponds to the
     range $0 < y < 1$, describing only a region $\M_1$ of the manifold $\M$
     of the solution (\ref{con-y}). The properties of the latter depend on
     the sign of $y_0$ \cite{br73}. In all cases, $y=0$ corresponds to a
     flat asymptotic, where $\phi \to \sqrt{6}y_0$, $|y_0| < 1$.

\medskip\noi
{\bf 3a:} $y_0 < 0$. The solution is defined in the range $0 < y <1/|y_0|$.
     At $y=1/|y_0|$, there is a naked attracting central singularity:
     $g_{tt}\to 0$, $r^2\to 0$, $\phi\to\infty$.

\medskip\noi
{\bf 3b:} $y_0 > 0$. The solution is defined in the range $y\in \R_+$.
     At $y\to\infty$, we find another flat spatial infinity, where
     $\phi\to \sqrt{6}/y_0$, $r^2\to\infty$ and $g_{tt}$ tends to a finite
     limit.  This is a {\sl wormhole solution\/}, found for the first time
     in \Ref{br73} and recently discussed by Barcelo and Visser \cite{viss}.

\medskip\noi
{\bf 3c:} $y_0=0$, $\phi=\sqrt{6}y$, $y\in \R_+$. In this case it is helpful
     to pass to the conventional coordinate $r$, substituting $y=m/(r-m)$.
     The solution
\bear
     ds^2 \eql (1-m/r)^2{dt^2} - \frac{dr^2}{(1-m/r)^2} -r^2d\Omega^2,
\nn
     \phi \eql \sqrt{6}m/(r-m)                              \label{con-bh}
\ear
     is the well-known BH with a conformal scalar field \cite{bbm70,bek74},
     which seems to violate the no-hair theorem. The infinite value of
     $\phi$ at the horizon $r=m$ does not make the metric singular since, as
     is easily verified, the energy-momentum tensor remains finite there.

     The whole case 3 belongs to variant III in the classification of \sect
     5, and the horizon in case 3c is situated in the region
     $\M_2 = \M \setminus \M_1$, where the action of the no-hair theorem
     cannot be extended.

     In case 3b, the second spatial infinity and even the wormhole throat
     ($y= 1/\sqrt{y_0}$) are situated in $\M_2$, illustrating the inferences
     of \sect 5.

     An important lesson follows, however, from case 2, where the mapping is
     type I by the same classification ($\M \map \oM$): there appears a
     minimum of $r(u)$ in the metric $g\mn$ (\ref{g-con}), and $r$ even
     blows up at large $u$.  This is connected with blowing up of the
     conformal factor $F$. Recall that, as mentioned in \sect 5, the absence
     of another spatial infinity is only guaranteed under the finiteness
     condition for the conformal factor in the whole range of the radial
     coordinate, including its boundary values: we see that this condition
     is indeed essential.

     The simple example of the conformal field thus illustrates the possible
     nontrivial consequences of conformal continuations. We shall see,
     however, that for most choices of STT and HOG one is guaranteed against
     such continuations.

\subsection{Conformal continuation conditions in scalar-tensor and
	high-order gravity}                                        

     Let us put for simplicity $D=4$ and consider possible conformal
     continuations of Einstein-frame solutions of STT and HOG
     due to transition to the Jordan frame.

     In STT (\ref{act-J}), such a continuation may occur at a zero of the
     function $f(\phi)$ in \eq (\ref{act-J}). If, at $\phi=\phi_0$, the
     function $f(\phi)$ has a simple zero, $f(\phi) = (\phi-\phi_0) \cdot
     O(1)$, then, in the transformation (\ref{g-wag}), (\ref{phi-wag}) for
     $D=4$ we have, without loss of generality,
\bear
     |\phi-\phi_0| \eql \e^{-\sqrt{2/3}\varphi}\cdot O(1),     \label{phi0}
\\
	F(\varphi) \eql \e^{\sqrt{2/3}\varphi}\cdot O(1)       \label{F0}
\ear
     as $\phi \to \phi_0$, so that $\varphi\to\infty$.

     In HOG (\ref{act-hog}) a continuation is possible if $f_R\to 0$ at some
     $R$. Then in the transformation (\ref{g-hog}), (\ref{phi-hog}) we
     obtain for $D=4$, without loss of generality,
\beq
	|f_R| \sim \e^{-\sqrt{2/3}\varphi}                     \label{fR0}
\eeq
     and again the relation (\ref{F0}). Thus the conformal factor has
     the same leading order behaviour in both theories.

     A conformal continuation of the metric (\ref{ds-d}) can obviously occur
     with the factor $F$ at some $\rho=\rho_0$ under the condition that the
     functions
\[
	F (\varphi)\,A(\rho), \cm F(\varphi)\,r^2(\rho)
\]
     have finite values at $\rho=\rho_0$. This means that $\rho=\rho_0$ is a
     generic regular sphere in the Jordan frame. Since this is a centre
     ($r\to 0$) in the Einstein frame, $\rho_0$ is finite, and we can put
     $\rho_0=0$ by a proper choice of the origin of $\rho$. Then \eqs
     (\ref{01d}) and (\ref{02d}) show that, in the leading order of
     magnitude,
\beq                                                         \label{con1}
     r^2(\rho) \sim A(\rho) \sim \sqrt{\rho} \sim \e^{-\sqrt{2/3}\varphi}.
\eeq
     Hence near $\rho=0$ the functions $r(\rho)$ and $A_\rho)$ may be
     represented by the expansions
\beq
     r = \rho^{1/4}(r_0 + r_1 \rho +...), \quad
     A = \rho^{1/2}(A_0 + A_1 \rho +...),                   \label{series}
\eeq
     where $r_0,\ r_1,\ldots,\ A_0,\ A_1,\ldots$ are constants. Substituting
     (\ref{series}) into the field equations, in particular, (\ref{00d}), we
     find that generically the potential $V(\varphi)$ behaves as
     $1/\sqrt{\rho}$, but it may happen that the leading order (or orders)
     vanish due to special relations between the expansion constants in
     (\ref{series}). One concludes, in general, that
\beq
     V(\varphi) \sim \rho^{-1/2 +n}, \cm n = 0, 1, 2, \ldots. \label{con-V}
\eeq

     Returning to \eqs (\ref{phi0}) and (\ref{F0}) and recalling that the
     potential function $U(\phi)$ in the STT (\ref{act-J}) is expressed in
     terms of $V$ as $U = V/F^2$ [see \eq(\ref{V-E})], we conclude that
     near $\phi_0$  \nq
\beq                                                          \label{con-U}
     F(\varphi) \sim |\phi-\phi_0|^{-1} \sim 1/\sqrt{\rho},
\quad
     U(\phi) \sim |\phi-\phi_0|^{1 + 2n},
\eeq
     where $n$ comes from (\ref{con-V}). We conclude that {\sl such a
     continuation is only possible when $U(\phi)$ has an odd-order zero at
     $\phi=\phi_0$.}

     For HOG we have the expression (\ref{V-hog}) for the potential $V$,
     which, for $D=4$, is rewritten as
\beq
      2V(\varphi) = (Rf_R - f)/f_R^2.                        \label{V-hog4}
\eeq
     In the case of interest, $f_R = 1/F(\varphi) \sim \sqrt{\rho}$, whereas
     $V(\varphi)$ either vanishes at $\rho=0$ or, at most, blows up as
     $1/\sqrt{\rho}$. This is only possible if $f(\phi_0)=0$. Thus {\sl a
     necessary condition for a continuation is that $f=f_R=0$
     simultaneously at some value of $R$.} Moreover, the requirement that
     $f(R)$ should be smooth at $R=R_0$ leaves the only opportunity
     $V \sim 1/\sqrt{\rho} \sim 1/f_R$; \eq (\ref{V-hog4}) then shows that
     $R_0 \ne 0$.

     Besides a generic sphere, a continuation may proceed through a horizon
     in the Jordan frame. In other words, in the metric
\beq                                                         \label{dsJ}
     	ds_{\rm J}^2 = F(\varphi)
     \biggl[A(\rho)dt^2 - \frac{d\rho^2}{A(\rho)} - r^2 d\Omega^2\biggr],
\eeq
     a certain value of $\rho$ (without loss of generality, $\rho=0$) may
     correspond to a horizon of order $k\geq 1$. This means that $\rho=0$ is
     a zero of order $k$ of the function $\oA(q)=AF$, where $q(\rho)$ is a
     new coordinate satisfying the condition $g_{tt}g_{qq}=-1$ in
     (\ref{dsJ}) (see the comment on the choice of the $\rho$ coordinate in
     \sect 2). As a result, we must have
\bearr
     \pm dq(\rho) = F(\varphi) d\rho;
\nnn
     \oA(q)= AF \sim (q-q_0)^k, \cm Fr^2 = O(1)                \label{horiz}
\ear
     where $q_0$ is the value of $q$ corresponding to $\rho=0$. As before,
     let us suppose that $F(\varphi) \sim \e^{\sqrt{2/3}\varphi}$ and
     $\varphi\to\infty$ as $\rho\to 0$.

     A substitution to \eq (\ref{01d}) leads, as before, to
     $r^2 \sim 1/F \sim \sqrt{\rho}$. A further substitution to (\ref{02d})
     then leaves two opportunities: (i) $A(\rho)\sim \sqrt{\rho}$ and (ii)
     $A(\rho) \sim \rho^{3/2}$.

     In the first case $AF$ tends to a finite limit, contrary to what was
     assumed (we simply return to the case of a generic regular sphere).

     In the second case, there can be a second-order horizon ($AF \sim \rho
     \sim (q-q_0)^2$). One can, however, show that, according to \eq
     (\ref{02d}), $A(\rho)<0$ as $\rho\to 0$, so this horizon is
     approached from a T region as $\rho\to +0$. If there is a static region
     at certain $\rho>0$, this means that, as $\rho$ decreases, $A(\rho)$
     changes its sign at some other horizon, say, $\rho = h>0$. Recalling
     the proof of Theorem 3, one can assert that $B(\rho) = A/r^2$ is a
     nondecreasing function at $\rho < h$. On the other hand, in the case
     under consideration one has $B \sim -r^4 \sim -\rho$ near $\rho=0$,
     i.e., a decreasing function. This contradiction shows that {\sl a
     continuation through a horizon in the Jordan frame is only possible
     when the whole region $\rho>0$ {\rm (the whole space in the Einstein
     frame)} is a T region.}

     In this case, as $\rho\to 0$, $V(\varphi)\sim 1/\sqrt{\rho}$. In STT
     this leads to $U(\phi)\sim \phi-\phi_0$. For HOG this is just the
     variant of $V$ admitted by (\ref{V-hog4}), and the requirement to
     $F(R)$ is the same as before: at $R=R_0\ne 0$, $f(R)$ should have
     at least a second-order zero.

     Summing up, we have the following two theorems and comment:

\Theorem{Theorem 6}
    {Consider \ssph\ solutions in STT (\ref{act-J}), $D=4$. Suppose that (a)
     $f(\phi) > 0$ at $\varphi <\varphi_0$; (b) $f(\varphi_0)=0$ but
     $df/d\phi(\phi_0) \ne 0$. Then the solution can be continued through
     the sphere where $\phi=\phi_0$ only if $U(\phi)$ has an odd-order zero
     at $\phi=\phi_0$.
     }

\Theorem{Theorem 7}
    {Consider \ssph\ solutions in HOG (\ref{act-hog}). Suppose that
     the function $f_R >0$ at $R < R_0$ and $f_R(R_0)=0$.
     Then the solution can be continued through the sphere where $R=R_0$
     only if $R_0\ne 0$ and $f(R)$ has an at least second order zero at
     $R=R_0$.
     }

\Theorem{Comment}
    {The sphere $\phi=\phi_0$ or $R=R_0$, admitting a continuation, can be
    (but not necessarily is) a horizon, and it is then double, only if the
    whole Einstein-frame solution represents a T region. In STT, under the
    conditions of Theorem 6, this can only happen if $U(\phi)$ has a simple
    zero at $\phi=\phi_0$.
    }

    One should stress that the conditions enumerated in Theorems 6 and 7 are
    only necessary for a possible continuation. It would be quite incorrect
    to think that any given solution to a theory satisfying these conditions
    may be continued in this way. This is perfectly well seen in the
    example of \sect 6.1: the potential $U(\phi)$ is zero identically, so
    the restriction of Theorem 6 is avoided, but a continuation
    actually takes place only for a special subfamily of the solutions,
    selected by a certain relation between the integration constants.

    On the other hand, Theorems 6 and 7 single out very narrow sets of
    theories from the class of STT and HOG considered. For all others, the
    Jordan-frame solutions obtained by conformal mappings from the Einstein
    frame are complete, and, in particular, Theorem 3 that determines the
    possible choice of global causal structures, is applicable.

\subsection*{Acknowledgements}

This work was supported in part by the Russian Foundation for Basic
Research. I am grateful to Oleg Zaslavskii, Mikhail Katanaev, Georgy Shikin
and Marek Biesiada for helpful discussions.

\end{document}